# Deep-space laser-ranging missions ASTROD (Astrodynamical Space Test of Relativity using Optical Devices) and ASTROD I for astrodynamics and astrometry

Wei-Tou Ni[1] and the ASTROD I ESA COSMIC VISION 2015-2025 TEAM

[1]Purple Mountain Observatory, Chinese Academy of Sciences, Nanjing, 210008, China
email: wtni@pmo.ac.cn

**Abstract.** Deep-space laser ranging will be ideal for testing relativistic gravity, and mapping the solar-system to an unprecedented accuracy. ASTROD (Astrodynamical Space Test of Relativity using Optical Devices) and ASTROD I are such missions. ASTROD I is a mission with a single spacecraft; it is the first step of ASTROD with 3 spacecraft. In this talk, after a brief review of ASTROD and ASTROD I, we concentrate of the precision of solar astrodynamics that can be achieved together with implications on astrometry and reference frame transformations. The precision planetary ephemeris derived from these missions together with second post-Newtonian test of relativistic gravity will serve as a foundation for future precise astrometry observations. Relativistic frameworks are discussed from these considerations.

**Keywords.** Astrodynamics, astrometry, tests of relativistic gravity, ASTROD, ASTROD I

1. Introduction

The improvement of the accuracy of Satellite Laser Ranging (SLR) and Lunar Laser Ranging (LLR) has great impact on the geodesy, reference frames and test of relativistic gravity among other things. At present, the accuracy for 2-color (2-wavelength) satellite laser ranging reaches 1 mm and that for 2-color lunar laser ranging are progressing towards a few millimeters. The present SLR and LLR are passive ranging; the intensity received at the photodetector is inversely proportional to the fourth power of the distance; for LLR, the received photons are scarce. However, for active ranging for which the laser light is received at the other end of ranging, the intensity received is inversely proportional to the second power of distance; even at interplanetary distance, the received photons would be abundant and deep-space laser ranging missions are feasible. The primary objective for the long-term ASTROD (Astrodynamical Space Test of Relativity using Optical Devices) concept is to maintain a minimum of 3 spacecraft in orbit within our solar system using laser interferometric ranging to ultimately test relativity and to search for gravitational waves. This is separated into 3 distinct stages, each with increasing orders of scientific benefits and engineering milestones. The first stage, ASTROD I, will comprise a single spacecraft in communication with ground stations using laser pulse ranging, orbiting the Sun at an average distance of approximately 0.6 AU. See figure 1 (left) for a schematic of the proposed orbit design.

The second phase, ASTROD II, also called ASTROD since it is the baseline of the general ASTROD mission concept, will consist of three spacecraft communicating with 1-2 W CW lasers. Two of these (S/C 1 and S/C 2) will be in separate solar orbits and a third (S/C) is to be situated at the Sun-Earth Lagrangian point L1 (or L2). It is this configuration which is capable of detecting gravitational effects induced by solar g modes. See figure 1 (right) for a depiction of the orbit design and configuration of the spacecraft 700 days after launch. The third and final stage, ASTROD III or Super ASTROD, will then explore the possibility of larger orbits in an effort to detect lower frequency primordial gravitational waves.

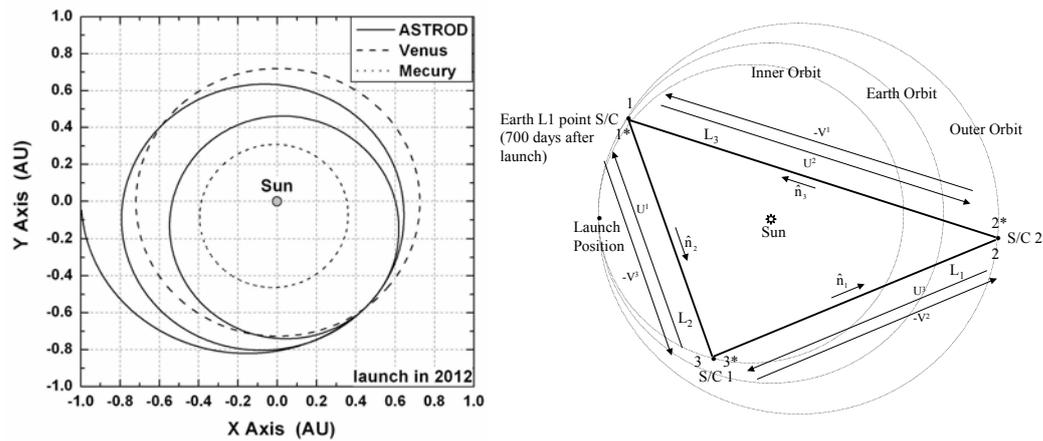

**Figure 1.** (left) A 2012 orbit in heliocentric ecliptic coordinate system for ASTROD I spacecraft; (right) A schematic ASTROD configuration (baseline ASTROD after 700 days from launch)

**2. ASTROD I**

The basic scheme of the ASTROD I space mission concept as proposed to ESA Cosmic Vision 2015-2025 is to use two-way laser pulse ranging between the ASTROD I spacecraft in solar orbit and deep space laser stations on Earth to improve the precision of solar-system dynamics, solar-system constants and ephemeris, to measure the relativistic gravity effects and test the fundamental laws of spacetime more precisely, and to improve the measurement of the time rate of change of the gravitational constant. A summary of ASTROD I goals is compiled in Table 1 (Appourchaux *et al.* 2007).

**Table 1:** Summary of the scientific objectives of the ASTROD I mission

| Effect/Quantity | Present accuracy | Projected accuracy |
|---|---|---|
| PPN parameter $\beta$ | $2 \times 10^{-4}$ | $3 \times 10^{-8}$ |
| PPN parameter $\gamma$ | $4.4 \times 10^{-5}$ | $3 \times 10^{-8}$ |
| Lense Thirring Effect | 0.1 | 0.1 |
| (dG/dt)/G | $10^{-12}$ yr$^{-1}$ | $3 \times 10^{-14}$ yr$^{-1}$ |
| Anomalous Pioneer acceleration $A_a$ | $(8.74 \pm 1.33) \times 10^{-10}$ m/s$^2$ | $0.7 \times 10^{-16}$ m/s$^2$ |
| Determination of solar quadrupole moment | $(1 - 3) \times 10^{-7}$ | $1 \times 10^{-9}$ |
| Determination of planetary masses and orbit parameters | (depends on object) | 1 - 3 orders better |
| Determination of asteroid masses and | (depends on object) | 2 - 3 orders better |

| densities | | |

To achieve these goals, the timing accuracy for ranging is required to be less than 3 ps and the drag-free acceleration noise at 100 μHz is required to be below $3 \times 10^{-14}$ m s$^{-2}$ Hz$^{-1/2}$. Figure 2 (left) shows the drag-free acceleration noise requirements for ASTROD I, the LTP, LISA and ASTROD for comparisons. To separate astrodynamical effect from relativistic-gravity test, the ASTROD I orbit is designed to reach the opposite side of the Sun at about 365 day and 680 day from launch. The apparent angles of the spacecraft during the two solar oppositions are shown in Fig. 2 (right) for the 2012 trajectory in Figure 1. The maximum one-way Shapiro time delays near the two solar oppositions are 0.1172 ms (at 365.3 day) and 0.1196 ms (at 679.1 days), respectively. The Shapiro time delay can then be measured precisely to have a good test of relativistic gravity.

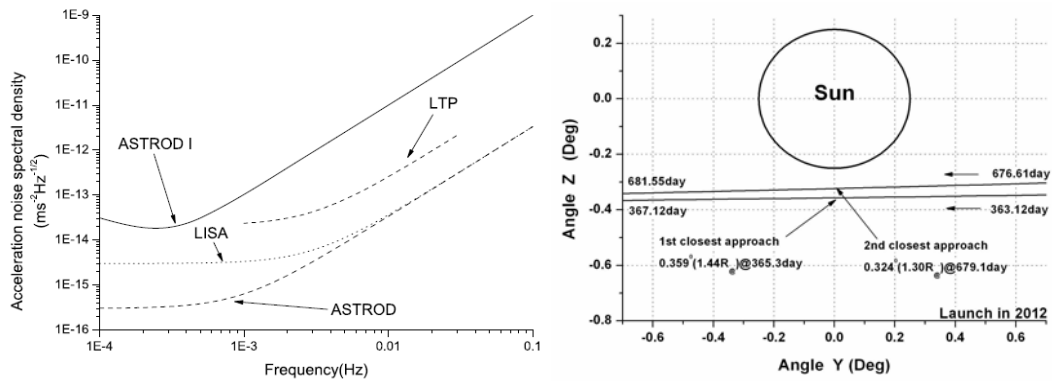

**Figure 2.** (left) A comparison of the target acceleration noise curves of ASTROD I, the LTP, LISA and ASTROD; (right) Apparent angles during the two solar oppositionsfor the ASTROD I orbit in figure 1.

Table 2 summarizes the astrodynamical and astrometric objectives of the ASTROD I and ASTROD missions:

**Table 2.** Summary of the astrodynamical and astrometric objectives of the ASTROD I and ASTROD missions

| Effect/Quantity | ASTROD I | ASTROD |
|---|---|---|
| Timing accuracy | 3 ps | 1 ps or better |
| Timing precision | 3 ps | 0.1 ps |
| Ranging accuracy | 0.9 mm | 0.3 mm or better |
| Ranging precision | 0.9 mm | 3-10 μm |
| Acceleration noise at 100 μHz | $3 \times 10^{-14}$ m s$^{-2}$ Hz$^{-1/2}$ | $3 \times 10^{-16}$ m s$^{-2}$ Hz$^{-1/2}$ |
| Derived angle accuracy | < 10 μas | 1 μas |
| Dynamical frame accuracy | < 100 μas | < 10 μas |

The angle accuracy depends on (i) derived angle accuracy from ranging of ASTROD I spacecraft and (ii) laser pointing measurement noise. With a ranging accuracy of 0.9 mm and acceleration noise as quoted, the derived azimuthal range accuracy would be much better than 1 m; this corresponds to 1.3 μas in angle at a distance of 1 AU. For laser pointing measurement accuracy, we assume it is below 10 μas. These give derived angle accuracy of below 10 μas in the table. For the dynamical frame accuracy, we conservatively down grade it by a factor of 10 to below 100 μas.

Just like LLR, ASTROD I (and ASTROD) will contribute to the determination of celestial reference frame, terrestrial frame and earth rotation. A dynamical realization of the International Celestial Reference System (ICRS) by the lunar orbit has been obtained from LLR to an uncertainty about 1 mas. A similar determination from ASTROD I orbit would leads to an uncertainty below 100 μas. The simultaneous data-fitting solution for the coordinates and velocities for the laser ranging stations will contribute to the realization of international terrestrial reference frame. Ranging to ASTROD I would contribute to the determination of nutation parameters due to dynamical stability of ASTROD I orbit.

## 3. ASTROD

ASTROD (ASTROD II) mission concept is to have two spacecraft in separate solar orbits carrying a payload of a proof mass, two telescopes, two 1–2 W lasers, a clock and a drag-free system, together with a similar spacecraft near Earth at one of the Lagrange points L1/L2 (Bec-Borsenberger *et al.* 2000; Ni 2002; Ni *et al.* 2004). The three spacecraft range coherently with one another using lasers to map the solar-system gravity, to test relativistic gravity, to observe solar g-mode oscillations, and to detect gravitational waves. Distances between spacecraft depend critically on solar-system gravity (including gravity induced by solar oscillations), underlying gravitational theory and incoming gravitational waves. A precise measurement of these distances as a function of time will determine these causes. After 2.5 years, the inner spacecraft completes 3 rounds, the outer spacecraft 2 rounds, and the L1/L2 spacecraft (Earth) 2.5 rounds. At this stage two spacecraft will be on the other side of the Sun, as viewed from the Earth, for conducting the Shapiro time delay experiment efficiently. The spacecraft configuration after 700 days from launch is shown in Figure 1 (right).

For a mission like ASTROD II within the next 10-20 years, a timing accuracy better than 1 ps (300 μm in terms of range) can be anticipated. In coherent interferometric ranging, timing events need to be generated by modulation/encoding technique or by superposing timing pulses on the CW laser light. The interference fringes serve as consecutive time marks. With timing events aggregated to a normal point using an orbit model, the precision can reach 30 μm in range. The effective range precision for parameter determination could be better, reaching 3-10 μm using orbit models. Since ASTROD range is typically of the order of 1-2 AU ($(1.5\text{-}3) \times 10^{11}$ m), a range precision of 3 μm will give a fractional precision of distance determination of $10^{-17}$. Therefore, the desired clock accuracy/stability is $10^{-17}$ over 1000 s travel time. Optical clocks with this accuracy/stability are under research development. Space optical clock is under development for the Galileo project. This development would pave the road for ASTROD to use optical clocks. With these anticipations, we list the astrodynamical and astrometric objectives in Table 2.

## 4. Relativistic frameworks

For the relativistic framework to test relativity and to do astrodynamics and astrometry, we use the scalar-tensor theory of gravity including intermediate range gravity which we used in obtaining the second post-Newtonian approximation (Xie *et al.* 2007). For the light deflection in the second post-Newtonian approximation in the solar field, please see Dong & Ni (2007). To extend to multi-bodies with multipoles in the solar system, we will need to extend our calculation following the work of Kopeikin & Vlasov (2004).

## Acknowledgements

We would like to thank the National Natural Science Foundation of China (Grant Nos 10475114 and 10778710) and the Foundation of Minor Planets of Purple Mountain Observatory for support.